\documentclass[11pt,oneside,letter]{article}
\usepackage[papersize={8.5in,11in}]{geometry}
\usepackage{authblk,amsmath,xcolor,suffix,paralist,amsfonts}
\usepackage[square,comma,numbers,sort&compress]{natbib}
\usepackage{hyperref}

\newcommand{\mnras}{Mon. Not. R. Astron. Soc.}
\newcommand{\apjl}{Astrophys. J. Letters}

\begin{document}

\title{Modified Gravity (MOG), the speed of gravitational radiation
  and the event GW170817/GRB170817A}

\author{M. A. Green$^{\dag}$, J. W. Moffat$^{\dag *}$ and V. T. Toth$^{\dag}$}

\affil{$^\dag$Perimeter Institute for Theoretical Physics, Waterloo,
  Ontario N2L 2Y5, Canada}
\affil{$^*$Department of Physics and Astronomy, University of
  Waterloo, Waterloo, Ontario N2L 3G1, Canada}

\maketitle

\begin{abstract}
  \noindent Modified gravity (MOG) is a covariant, relativistic,
  alternative gravitational theory whose field equations are derived
  from an action that supplements the spacetime metric tensor with
  vector and scalar fields.  Both gravitational (spin 2) and
  electromagnetic waves travel on null geodesics of the theory's one
  metric.  Despite a recent claim to the contrary, MOG satisfies the
  weak equivalence principle and is consistent with observations of
  the neutron star merger and gamma ray burster event
  GW170817/ GRB170817A.
  \end{abstract}

\noindent The nearly simultaneous detection of gravitational wave and
electromagnetic \mbox{$\gamma$-ray} signals from the merging neutron
star event GW170817/GRB170817A imposes stringent limits on violation
of the weak equivalence principle (WEP) \cite{Abbot2017a,
  2017arXiv171005860W}.  Together with constraints on gravitational
Cherenkov energy loss, this rules out, or severely constrains, many of
the modified gravity theories\,---\,scalar-tensor, vector-tensor,
bimetric\,---\,that have been proposed to avoid the need for dark
matter and/or dark energy \cite{Chesler2017,Sagi,2017arXiv171005877C,
  2017arXiv171005893S, Ezquiaga2017, Boranetal2017, Bakeretal2017}.

Boran et al (BDKW) \cite{Boranetal2017} focus on models that
``dispense with the dark matter paradigm and reproduce Modified
Newtonian Dynamics (MOND) like behavior in the non-relativistic
limit''.  BDKW call such models ``Dark Matter Emulators'', and define
them explicitly as modified gravity theories in which:
\begin{enumerate}
\item Ordinary matter couples to the metric $\tilde{g}_{\mu\nu}$
  ($\tilde{g}$ denotes the ``disformally transformed metric'') that
  would be produced by general relativity with dark matter; and
\item Gravitational waves couple to the metric $g_{\mu\nu}$ produced
  by general relativity without dark matter.
\end{enumerate}
Citing TeVeS \cite{Bekenstein,Sagi} and MOG (Moffat's modified gravity, also known as
Scalar-Tensor-Vector-Gravity (STVG)~\cite{Moffat2006a}) as examples of
``Dark Matter Emulators'', BDKW argue that ``the whole class'' of such
models predict large Shapiro delays incompatible with the
GW170817/GRB170817A observations.  In fact, MOG is not a bimetric
theory and does not match the above-specified criteria for ``Dark Matter
Emulators''.  MOG has only one metric, both photons and gravitons travel
on null geodesics of this metric, and their nearly simultaneous
detection is what one would expect with MOG.

Indeed, MOG was not designed to reproduce MOND in the nonrelativistic
limit.  MOG has been shown to fit galaxy dynamics, cluster dynamics
and early universe cosmology data (see, e.g.,
\cite{MoffatRahvar2013,MoffatRahvar2014}) without detectable dark
matter in the present universe.  By contrast, MOND is unable to fit
cluster observations (without dark matter) or cosmology observations.

The MOG theory is based on three gravitational fields: (1) the Einstein
metric $g_{\mu\nu}$ corresponding, in particle physics language, to a
spin 2 massless graviton, (2) a scalar field $G=G_N(1+\alpha)$
corresponding to a spin 0 massless graviton, (3) a vector field
$\phi_\mu$ corresponding to a spin 1 (repulsive) massive graviton.
There is also an effective spin 0 scalar field $\mu$\,---\,the mass of
the vector field $\phi_\mu$.

MOG is a classical Lagrangian field theory, with a fully covariant
action $S$ given in terms of the metric $g_{\mu\nu}$, cosmological
constant $\Lambda$, vector field $\phi^\mu$, and scalar fields $G$ and
$\mu$~\cite{Moffat2006a}:\footnote{The scalar field $\omega(x)$ introduced in
  \cite{Moffat2006a} is here made constant: $\omega=1$.}
\begin{align}
S=S_G+S_\phi+S_S+S_M,
\end{align}
where $S_M$ is the matter action and
\begin{align}
S_G&=\frac{1}{16\pi}\int d^4x\sqrt{-g}\left[\frac{1}{G}(R+2\Lambda)\right],\\
S_\phi&=\int d^4x\sqrt{-g}\left[-\frac{1}{4}B^{\mu\nu}B_{\mu\nu}+V_\phi\right],\\
S_S&=\int d^4x\sqrt{-g}\left[\frac{1}{G^3}\left(\frac{1}{2}g^{\mu\nu}\partial_\mu G\partial_\nu G-V_G\right)+\frac{1}{\mu^2G}\left(\frac{1}{2}g^{\mu\nu}\partial_\mu\mu\partial_\nu\mu-V_\mu\right)\right],
\end{align}
with
$B_{\mu\nu}=\partial_\mu\phi_\nu-\partial_\nu\phi_\mu$.  Note that we choose units such that
$c=1$, use the metric signature $[+,-,-,-]$, and define the sign of
the Ricci tensor such that
$R_{\mu\nu}=\partial_\alpha\Gamma^\alpha_{\mu\nu}-...$\,.

The matter stress-energy tensor is obtained by varying the
matter action $S_M$ with respect to the metric:
\begin{align}
  T^{\mu\nu}_M=-2(-g)^{-1/2}\delta S_M/\delta g_{\mu\nu}\,.
\end{align}
Varying $S_\phi+S_S$ with respect to the metric yields
\begin{align}
  T^{\mu\nu}_{\rm MOG}=-2(-g)^{-1/2}[\delta S_\phi/\delta
  g_{\mu\nu}+\delta S_S/\delta g_{\mu\nu}]\,.
\end{align}
Combining these gives the total stress-energy tensor
\begin{align}
  T^{\mu\nu}=T^{\mu\nu}_{\rm M}+T^{\mu\nu}_{\rm MOG}\,.
\end{align}

Varying the complete action $S$ with respect to $g_{\mu\nu}$ gives the
field equations:
\begin{align}
  G_{\mu\nu}-\Lambda g_{\mu\nu}+Q_{\mu\nu}=8\pi G\left(T_{\mu\nu}^M
  +T_{\mu\nu}^{\rm MOG}\right)\,,\label{eq:MOGE}
\end{align}
where $G_{\mu\nu}=R_{\mu\nu}-\frac{1}{2} g_{\mu\nu} R$
is the Einstein tensor and
\begin{align}
  Q_{\mu\nu}=\frac{2}{G^2}(\nabla^\alpha G \nabla_\alpha G\,g_{\mu\nu}
  - \nabla_\mu G\nabla_\nu G) - \frac{1}{G}(\Box G\,g_{\mu\nu}
  - \nabla_\mu\nabla_\nu G)\label{eq:Q}
\end{align}
is a boundary term resulting from the the presence of second derivatives
of $g_{\mu\nu}$ in $R$ in $S_G$.
Combining the Bianchi identities, $\nabla_\nu G^{\mu\nu}=0$, with the
field equations (\ref{eq:MOGE}) yields the conservation law
\begin{align}
  \nabla_\nu T^{\mu\nu}+\frac{1}{G}\nabla_\nu G\,T^{\mu\nu} -
  \frac{1}{8\pi G}\nabla_\nu Q^{\mu\nu}=0 \,.\label{eq:Conservation}
\end{align}

It is a key premise of MOG that all baryonic matter possesses, in
proportion to its mass $M$, positive gravitational charge:
$Q_g=\kappa\,M$.  This charge serves as the source of the vector field
$\phi^\mu$.  Moreover, $\kappa=\sqrt{G-G_N}=\sqrt{\alpha\,G_N}$, where
$G_N$ is Newton's gravitational constant and
$\alpha=(G-G_N)/G_N\ge 0$.  Variation of $S_M$ with respect to the
vector field $\phi^\mu$ then yields the MOG 4-current
$J_\mu=-(-g)^{-1/2}\delta S_M/\delta \phi^\mu$.  Variation of $S_\phi$
with respect to $\phi_\mu$ yields
\begin{equation}
\frac{1}{\sqrt{-g}}\partial_\mu\biggl(\sqrt{-g}B^{\mu\nu}\biggr)+\frac{\partial V_\phi}{\partial\phi_\nu}=-J^\nu,
\end{equation}
where we choose $V_\phi=\frac{1}{2}\mu^2\phi^\mu\phi_\mu$. In the
case of a perfect fluid with $T_\mu^\mu>0$ and 4-velocity $u^\mu$, the
4-current is given by $J_\mu=\kappa\,T_{\mu\nu}u^\nu$.  It is shown in
\cite{Roshan} that, with the assumption $\nabla_\mu J^\mu=0$,
(\ref{eq:Conservation}) reduces to
\begin{align}
  \nabla_\nu T_M^{\mu\nu}=B_\nu^{~\mu} J^\nu\,.\label{eq:Mcons}
\end{align}

The dimensionless scalar field $\alpha$ is zero for general
relativity.  For $\alpha > 0$ the strength of gravity is enhanced on
cosmological scales ($G=G_N(1+\alpha)$) and emulates dark matter in
lensing even though the matter source is baryons
\cite{2012arXiv1204.2985M}.  From fits to galaxy rotation curves of
galaxies and clusters, without introducing dark matter, $\alpha=8.89$
and $\mu=0.042 {\rm kpc}^{-1}$ (see
\cite{MoffatRahvar2013,MoffatRahvar2014}).  We note that
$\mu^{-1}=24$~kpc corresponds to a mass
$m_\phi=2.8\times 10^{-28}$~eV\,---\,of order the experimental bound
on the photon mass.  Due to the smallness of the vector particle mass,
the three gravitons $g_{\mu\nu}$, $G=G_N(1+\alpha)$ and $\phi_\mu$
move at the speed of light.  So, in MOG gravitational waves
(gravitons) move at the speed of light as do photons.

The MOG equation of motion of a test particle with mass $m>0$, and
free of baryonic forces, is \cite{Moffat2006a,Roshan}:
\begin{align}
  m\left(\frac{du^\mu}{ds}+\Gamma^\mu_{\alpha\beta}u^\alpha
  u^\beta\right)=q_g B^\mu_\alpha u^\alpha,\label{eq:EqMoMassive}
\end{align}
where $u^\mu=dx^\mu/ds$ and $q_g=\kappa\,m$ is the MOG charge of
the test particle \cite{Moffat2006a,Roshan}.  Cancelling
the mass $m$ shows that all particles (bodies) follow spacetime paths
independent of their constitution (satisfying the weak equivalence
principle).  However, due to the Lorentz-like acceleration that
applies universally to massive particles, massive test particles do
not free fall on geodesics.

For electromagnetic radiation, $m_\gamma=0$ implies $q_g=0$.
Photons (and other massless particles) thus follow null geodesics:
\begin{align}
\label{eq:EqMoMassless}
  k^\mu\nabla_\mu k^\nu=0,
\end{align}
where $k^\mu$ is the 4-momentum null vector and $k^2=g_{\mu\nu}k^\mu k^\nu=0$.
Gravitational radiation (gravitons) follows the same null geodesics
given by (\ref{eq:EqMoMassless}).

The problem with all the relativistic extensions of Milgrom's MOND
published to-date is that they are based on a disformal bimetric
geometry. The speed of light is subluminal, causing photons and
gravitons to move along different trajectories. This causes problems with
Cherenkov radiation and contradictions with cosmic rays and the
GW170817 data \cite{2017arXiv171005877C, 2017arXiv171005893S,
  Ezquiaga2017, Boranetal2017, Bakeretal2017}. Since all the
relativistic extensions of MOND are now ruled out, the nonrelativistic
MOND model seems left in an untenable position.  Other more
complicated Horndeski scalar-tensor theories \cite{Horndeski} are also
eliminated by the new WEP constraint.

In contrast with TeVeS type theories, MOG has only one metric
$g_{\mu\nu}$.  Photons and gravitons move along the same null geodesic
paths, as in general relativity, and will arrive at a destination at the
same time.  Since the theory is based on an action principle, the
field equations are generally covariant and the MOG vector field
$\phi_\mu$ is not a classical time-like unit vector as in TeVeS type
theories.  Neither does the Jordan-Brans-Dicke scalar-tensor sector of
MOG contradict the GW170917/GRB170817 data at the cosmological scale
\cite{Bakeretal2017}.

The BDKW authors have incorrectly associated the MOG theory with TeVeS
type theories in which $G$ is not enhanced to emulate dark matter.  (In
TeVeS type theories $G_N$ is renormalized by a small amount that
cannot emulate dark matter.)  At this stage, it appears that only MOG
is able to eliminate the need for dark matter and fit data including
GW170817/GRB170817A.

\section*{Acknowledgments}

This research was supported in part by Perimeter Institute for
Theoretical Physics. Research at Perimeter Institute is supported by
the Government of Canada through the Department of Innovation, Science
and Economic Development Canada and by the Province of
Ontario through the Ministry of Research, Innovation and Science.

\bibliographystyle{hunsrtnat.bst}

\end{document}